\documentclass[twocolumn,floatfix,prb,aps,showpacs]{revtex4}
\usepackage{graphicx,color,amsmath,amssymb}
\begin{document}
\title{State Counting for Excited Bands of the Fractional Quantum Hall Effect: Exclusion Rules for Bound Excitons}
\author{Ajit C. Balram,$^1$  Arkadiusz W\'ojs,$^2$ and Jainendra K. Jain,$^1$}
\affiliation{
   $^{1}$Department of Physics, 
   104 Davey Lab, 
   Pennsylvania State University, 
   University Park PA, 16802}
\affiliation{
   $^{2}$Institute of Physics, 
   Wroclaw University of Technology,
   50-370 Wroclaw, Poland}

\date{\today}

\begin{abstract}
Exact diagonalization studies have revealed that the energy spectrum of interacting electrons in the lowest Landau level splits, non-perturbatively, into bands, which is responsible for the fascinating phenomenology of this system. The theory of nearly free composite fermions has been shown to be valid for the lowest band, and thus to capture the low temperature physics, but it over-predicts the number of states for the excited bands. We explain the state counting of higher bands in terms of composite fermions with an infinitely strong short range interaction between an excited composite-fermion particle and the hole it leaves behind. This interaction, the form of which we derive from the microscopic composite fermion theory, eliminates configurations containing certain tightly bound composite-fermion excitons.  With this modification, the composite-fermion theory reproduces,for all well-defined excited bands seen in exact diagonalization studies, an exact counting for $\nu>1/3$, and an almost exact counting for $\nu\leq 1/3$. The resulting insight clarifies that the corrections to the nearly free composite fermion theory are not thermodynamically significant at sufficiently low temperatures, thus providing a microscopic explanation for why it has proved successful for the analysis of the various properties of the composite-fermion Fermi sea. 
\end{abstract}

\maketitle

\section{Introduction}

The eigenspectrum of interacting particles contains fundamental information about its physics. A non-perturbative restructuring of the low energy spectrum by the interaction sometimes signals the formation of complex quasiparticles that are in general distinct from the original particles and may even carry unusual quantum numbers \cite{Anderson}. An important example is that of interacting electrons in the lowest Landau level (LL). The most striking aspect of its exact Coulomb spectrum is that it splits, nonperturbatively, into mini-bands \cite{DevJainPRL92}, referred below to as ``$\Lambda$ bands."  The $\Lambda$ bands owe their existence entirely to interactions, and their formation lies at the heart of many phenomena, most notably the fractional quantum Hall effect (FQHE) \cite{TsuiStormerGossard82}. Therefore, an understanding of the physical origin of these bands and their structure as a function of the filling factor $\nu$ is a central problem of the FQHE.  

The non-perturbative physics of bands within the lowest LL is understood as a consequence of the formation of composite fermions. The composite fermion (CF) theory \cite{Jain1989,Jain_book,Quinn} postulates a mapping of the problem of interacting electrons at filling factor $\nu$ into that of composite fermions at filling factor $\nu^*$, the two related by $\nu=\nu^*/(2p\nu^*\pm 1)$ where $p$ is a positive integer. 
The underlying physics is  that electrons capture $2p$ vortices each as a result of the repulsive interaction; this, in turn, produces a dynamics, due to the Berry phases produced by the bound vortices, as though composite fermions experienced a reduced magnetic field given by $B^*=B-2p\rho\phi_0$, where $\phi_0=hc/e$ is the flux quantum. If one neglects the interactions between composite fermions, then the CF theory predicts that the $\Lambda$ bands at $\nu$ have a one to one mapping to the familiar ``Landau bands" at $\nu^*$
\begin{equation}
{\rm free}\;{\rm CF}\; {\rm model}:\;\Lambda\;{\rm bands}\; {\rm at}\; \nu \Leftrightarrow {\rm Landau}\;{\rm bands}\; {\rm at}\; \nu^*
\nonumber
\end{equation}
Here, the term ``Landau bands" refers to the bands in the {\em many particle spectrum}, to be distinguished from Landau {\em levels}, which are the energy levels of a single particle; of course, all eigenstates of the Landau bands can be enumerated straightforwardly from the knowledge of the single particle Landau levels.  Should this mapping be confirmed, one can interpret the $\Lambda$ bands as arising from single-particle ``$\Lambda$ levels" ($\Lambda$Ls) of composite fermions, which are analogous to the electron LLs at $\nu^*$, thus enabling a ``single-particle-like" explanation of the FQHE. It should be stressed that, unlike LLs, the $\Lambda$Ls are a nontrivial emergent concept that cannot be derived at a single particle level. The only way to confirm their reality is to compare the above correspondence between the exact many particle spectra of interacting electrons at $\nu$ and the Landau bands at $\nu^*$.  In other words, while we construct many body states of electrons starting from single particle LLs, for composite fermions we must carry out the inverse program: we only have the many body states to begin with, from which we must deduce the existence of $\Lambda$ levels. A single particle derivation of $\Lambda$ levels is not possible, because composite fermions themselves are collective entities defined only inside the background of other composite fermions. 

The important question, of course, is: How well does the mapping work?  Previous studies have shown that the {\em lowest} $\Lambda$ band at $\nu$ does indeed resemble the lowest Landau band at $\nu^*$. 
The correct account of the states below the gap is fortunately sufficient for the explanation of many phenomena occurring at low temperatures. In particular, the FQHE at the Jain fractions $\nu=n/(2pn\pm 1)$ is explained as the IQHE of composite fermions, with the ground state represented by $n$ filled $\Lambda$Ls of composite fermions.

The mapping, however, breaks down qualitatively for excited bands. As shown by Wu and Jain \cite{WuJain95}, the analogy to free fermions at $\nu^*$ predicts spurious ``unphysical" states in higher bands, i.e. states without counterparts in the $\Lambda$ bands seen in exact diagonalization. The number of such unphysical states grows with increasing energy. 

Why should we be concerned about excited bands? We would of course like to gain as complete an understanding of this quintessential many-body system as possible, but there are also at least two experimental motivations behind our current study. One, while many phenomena are governed by the low lying states, high energy excitations can also be probed, e.g., by light scattering. In particular, the existence of higher $\Lambda$ bands has been invoked for an explanation \cite{Majumder09} of the high energy collective modes observed in Raman experiments of Hirjibehedin {\em et al.} \cite{Hirjibehedin05} and Rhone {\em et al.}\cite{Rhone11} Two, the distinction between the lowest and the higher $\Lambda$ bands vanishes in the limit of $\nu=1/2$; here the gaps separating the $\Lambda$ bands collapse and {\em all} $\Lambda$Ls participate in the low energy manifold. One may therefore question the applicability of the model of nearly free composite fermions  for the thermodynamics of the CF Fermi sea state at $\nu=1/2$ \cite{HalperinLeeRead1993}.  Notwithstanding, the model of nearly free composite fermions has proved successful in understanding experiments. (The Chern Simons theory of the 1/2 state developed by Halperin, Lee and Read \cite{HalperinLeeRead1993} also does not contain the physics of the missing states.) A number of experimental results, such as the Shubnikov de Haas oscillations \cite{Du94,Leadley}, spin polarization \cite{Melinte,Freytag02}, valley polarization \cite{ShayeganValley}, commensurability oscillations \cite{Willett,Kang,Goldman,Smet,Shayegan1,Shayegan2}, thermopower \cite{Ying} and cyclotron resonances \cite{Kukushkin1,Kukushkin2} of composite fermions have been successfully analyzed in terms of a Fermi sea of nearly free composite fermions. This suggests that the model of nearly free composite fermions ought to be valid to at least some extent, but that is not obvious, {\em a priori}, in light of the above discussion of the missing states. 

This raises several conceptual questions. How do we understand the excited $\Lambda$ bands of the exact spectra? 
Do {\em empty} $\Lambda$Ls really exist above the CF Fermi level? (The existence of filled $\Lambda$ levels has been conclusively demonstrated by a variety of means, but relatively few studies have probed the empty $\Lambda$ levels \cite{Rhone11}.) How do we reconcile the experimental successes of the nearly free-CF model with the theoretical discrepancies mentioned above? We answer these questions in this paper.

\begin{figure*}[t]
\begin{center}
\includegraphics[width=1.0\textwidth]{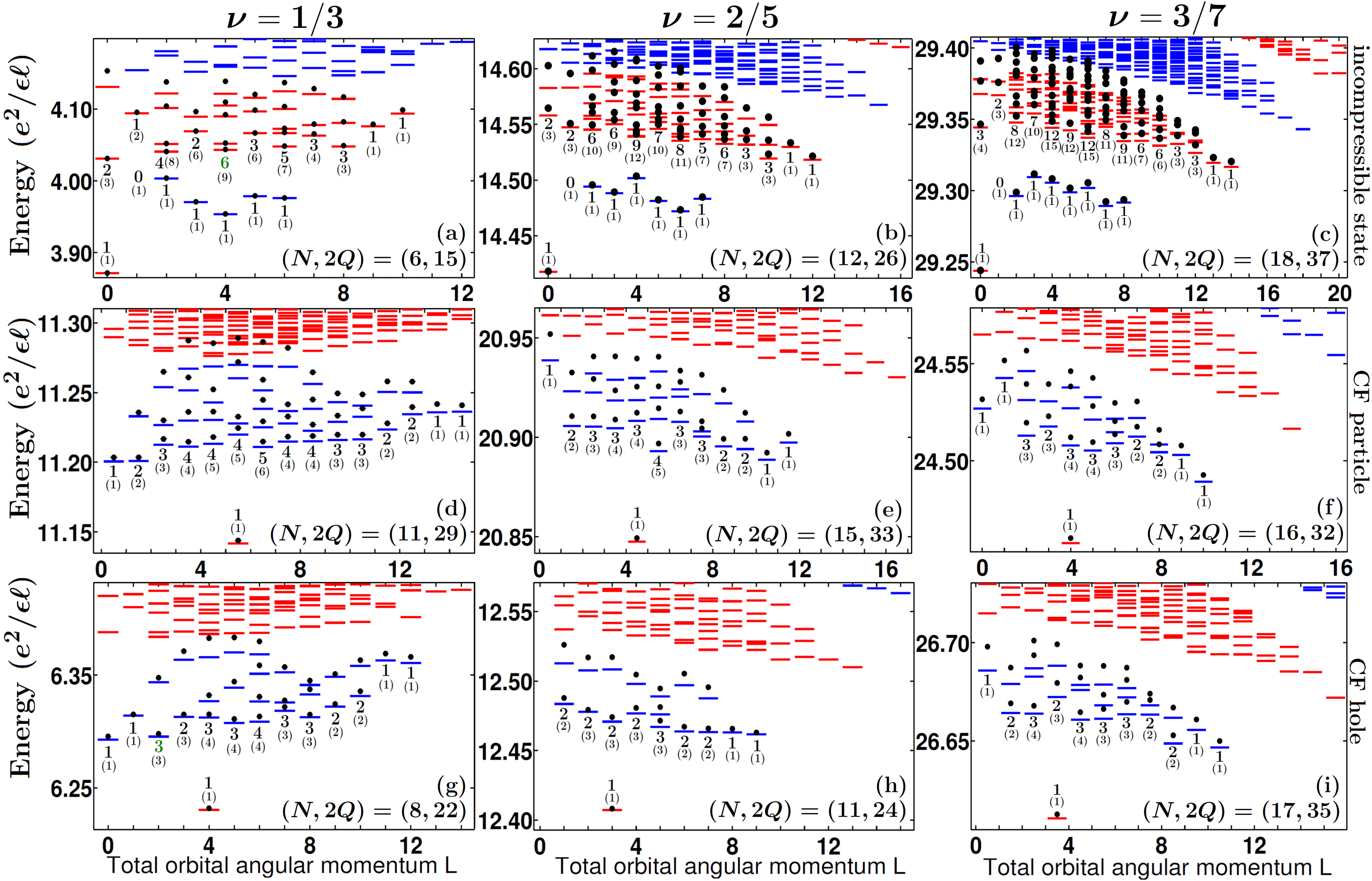}
\caption{(color online) Emergent ``$\Lambda$ bands" in the FQHE spectra. The exact Coulomb energies are shown by dashes, obtained in the spherical geometry for $N$ electrons at total flux $2Q (hc/e)$. The left, middle and right panels correspond to filling factors $\nu=$ 1/3, 2/5, and 3/7, respectively. The top panel shows the spectra for the incompressible state, while the middle and bottom panels show spectra of systems that contain a single CF particle or CF hole in the ground state. The energy is quoted in units of $e^2/\epsilon \ell$, where $\ell=\sqrt{\hbar c/eB}$ is the magnetic length and $\epsilon$ is the dielectric constant of the host medium. Alternate bands are colored in red and blue. The integers in brackets give the number of $L$ multiplets in the Landau bands of free fermions at $(N,2Q^*)$ with $Q^*=Q-N+1$. The integers outside the bracket show the number of multiplets that remain after eliminating the $L=1$ CF exciton; these agree with the actual number of multiplets except in two cases in panels (a) and (g), marked in green. The dots are the CF spectra obtained by CF diagonalization.
}
\label{spectra}
\end{center}
\end{figure*}

Our principal result is to show that the the concept of empty $\Lambda$ levels above the CF Fermi level is perfectly sensible, and that the excited $\Lambda$ bands can be constructed faithfully within the CF theory, provided we impose a strong short-range repulsion between a CF particle (CFP) and a CF hole (CFH) to eliminate certain tightly bound CF excitons. Here ``CFP" refers to a composite fermion in an otherwise empty $\Lambda$L, and ``CFH" to a missing composite fermion in an otherwise full $\Lambda$L. Elimination of these CF excitons produces an excellent account of the $\Lambda$ bands as seen in exact diagonalization studies.

This understanding clarifies that the number of eliminated states is thermodynamically insignificant at low temperatures, when the density of CFPs and CFHs is dilute.  Our calculations also show, surprisingly, that the residual interactions between composite fermions becomes weaker as we approach the CF Fermi sea at the half filled LL. These findings provide a microscopic justification for the adequacy of the nearly-free CF model for the low temperature thermodynamics of the FQHE states as well as of the CF Fermi sea.

The picture that has emerged reveals two kinds of interactions between composite fermions. One is a weak interaction that causes $\Lambda$ band broadening but does not change the number of states in each $\Lambda$ band. Accurate methods have been developed to treat that interaction. The other interaction, the one identified in this paper, is infinitely strong and eliminates some states by sending them to infinity -- this interaction acts as a constraint that projects out certain configurations, thereby altering the structure of the $\Lambda$ bands. The latter interaction will be seen to have a strange form, in that it acts only between a CFP and a CFH occupying different $\Lambda$Ls.  With these two interactions, the CF theory provides an excellent quantitative description of the $\Lambda$ bands wherever they can be clearly identified in exact Coulomb spectra (we have tested three lowest bands at 1/3, 2/5 and 3/7, and two lowest bands at filling factors in between).

The plan of the paper is as follows. In the next section, we show results that illustrate the discrepancy between the ``naive" prediction from the non-interacting CF theory and exact diagonalization. Section III contains a postulate for the strong interaction between composite fermions. This postulate is shown to be qualitatively valid in Section IV and quantitatively accurate in Section V. The paper is concluded with some remarks on the relevance of our results to experiments. We assume fully spin polarized electrons throughout this study, as our goal is to establish the principles governing the counting of the levels in the excited bands. 

\section{The puzzle of ``Missing states" in the excitation spectrum}

\begin{figure*}[h]
\centering
\includegraphics[width=1.0\textwidth]{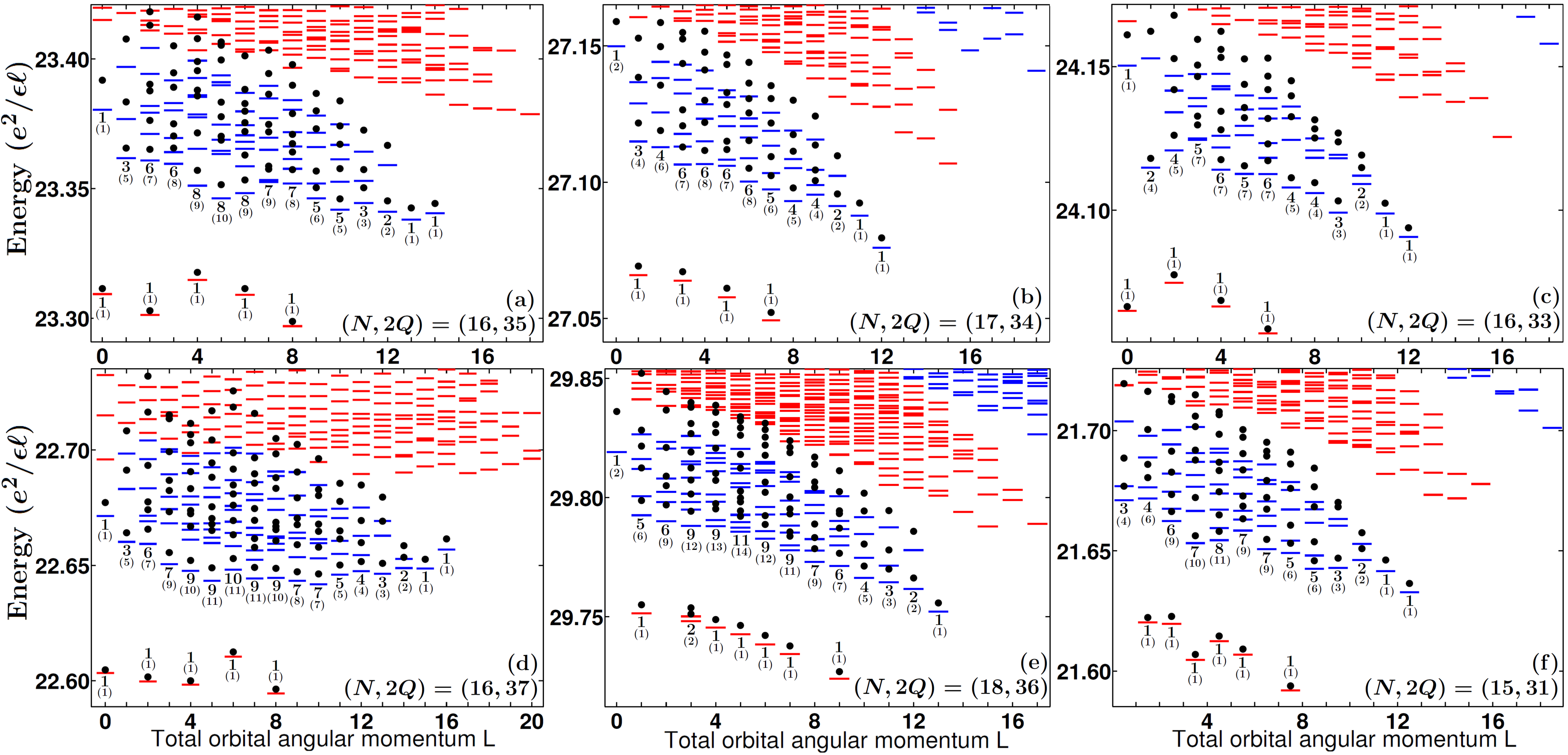}
\caption{Exact Coulomb spectra are shown for several systems, whose ground band corresponds to several QPs or QHs: (a) two QPs of 2/5; (b) two QPs of 3/7; (c) two QHs of 3/7; (d) two QHs of 2/5;  (e) three QPs of 3/7; (f) three QHs of 3/7. The energies levels are shown by dashes and alternate $\Lambda$ bands are colored in red and blue for ease of identification of different bands.  The black dots show the CF energies obtained from CF Diagonalization in the basis derived from all states up to the first excited Landau band. The integers in the brackets give the number of $L$ multiplets in the Landau bands of free fermions at $(N,2Q^{*})$, where $2Q^{*}=2Q-2(N-1)$. The integers outside the brackets show the number of multiplets that remain after excluding the single excitons listed in Eq.~\ref{exclusion}, and agree with the actual numbers of multiplets in all cases.}
\label{general}
\end{figure*}

\begin{figure*}[h]
\centering
\includegraphics[width=0.95\textwidth]{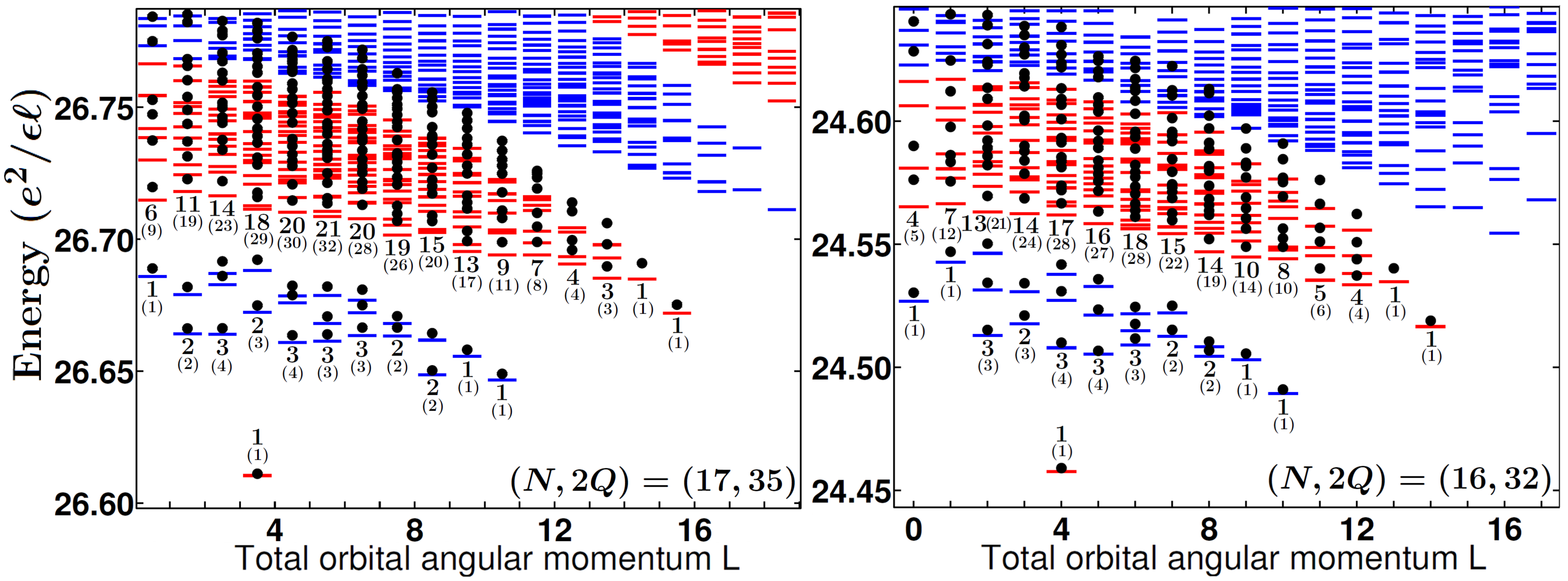}
\caption{Exact Coulomb spectra for $(N,2Q)=(17,35)$ corresponding to a single QH at 3/7 (left panel) and $(16,32)$ corresponding to a QP at 3/7 (right panel); energies levels are shown by dashes and alternate $\Lambda$ bands are colored in red and blue. The black dots show the CF energies obtained from CF Diagonalization in the basis derived from all states up to the second excited Landau band. The integers in the brackets give the number of $L$ multiplets in the Landau bands of free fermions at $(N,2Q^{*})$, where $2Q^{*}=2Q-2(N-1)$. The integers outside the brackets show the number of multiplets that remain after excluding the single excitons listed in Eq.~\ref{exclusion}, and agree with the actual numbers of multiplets in all cases.}
\label{second}
\end{figure*}

The essential puzzle can be seen in Figs. \ref{spectra}, \ref{general} and \ref{second}. These figures show Coulomb spectra of interacting electrons in the lowest LL in the spherical geometry \cite{Haldane1983} as a function of $N$, the number of electrons, and $2Q$, the flux through the surface of the sphere (in units of $\phi_0=hc/e$). $L$ is the total orbital angular momentum. The eigenenergies are obtained exactly by a brute force diagonalization of the Coulomb Hamiltonian. The energies are shown in units of $e^2/\epsilon \ell$, where $\ell=\sqrt{\hbar c/eB}$ is the magnetic length and $\epsilon$ is the dielectric constant of the background material.

The top row of Fig.~\ref{spectra} depicts several exact Coulomb spectra for the incompressible states at 1/3, 2/5 and 3/7; the middle and bottom rows show spectra for systems each of whose ground states contains a single CFP or a single CFH. (Note that in all cases, the {\em excited} states contain one or more CFPs and CFHs.) In each case, we have taken the largest $N$ for which the relevant bands are well defined; for larger $N$ the bands do exist, but begin to overlap, making it difficult to identify them clearly as would be needed for an unambiguous counting of states within each band. Fig.~\ref{general} shows spectra for systems for which the ground band contains several CFPs or CFHs, as indicated in the figure caption. In some cases, the second excited bands are also identifiable; two such examples are shown in Fig.~\ref{second} where the ground band contains a single CFP or CFH in the 3/7 state.

In all cases, the spectrum contains $\Lambda$ bands that are identifiable by inspection. For clarity, we have shown the states of successive bands with different colors. In the absence of interelectron interaction, {\em all} states would degenerate (because all electrons are occupying the lowest LL), producing a degenerate Landau band. It is expected that the Coulomb interaction will broaden the Landau band, but there is no {\em a priori} reason to suspect that it would split into $\Lambda$ bands. The emergence of $\Lambda$ bands is a nonperturbative effect, persisting as we reduce the strength of the Coulomb interaction, say by taking the dielectric function $\epsilon$ to be very large. 

The CF theory explains the formation of $\Lambda$ bands within the lowest Landau band as follows. It postulates the formation of composite fermions, bound states of electrons and $2p$ vortices, which experience an effective magnetic field $B^*=B-2p\rho\phi_0$. It further postulates that composite fermions can be treated, to a first approximation, as noninteracting. The FQHE problem is thus mapped into the problem of noninteracting fermions at $B^*$. Translating into the spherical geometry in the standard manner, the problem of interacting electrons at $2Q$ is mapped into that of noninteracting fermions at an effective flux $2Q^*=2Q-2p(N-1)$. We allow fermions at $2Q^*$ to occupy higher LLs. The $\Lambda$ bands at $2Q$ are thus postulated to be related to the Landau bands at $2Q^*$.

\begin{figure}
\begin{center}
\includegraphics[width=0.45\textwidth]{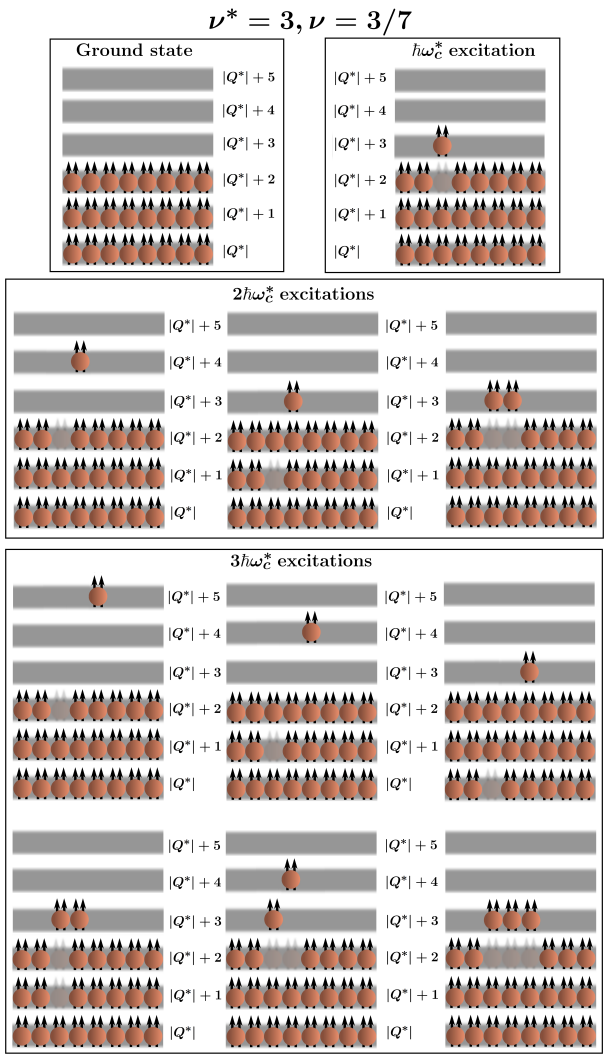}
\caption{(color online) This figure shows all configurations up to energy less than or equal to 3 units of CF cyclotron energy ($\hbar\omega_c^*$) at $\nu^*=3$ (which corresponds to the electron filling $\nu=3/7$). The angular momentum of each $\Lambda$ level is shown; it is $|Q^*|$ for the lowest $\Lambda$ level and increases by one unit for each successive $\Lambda$ level.
}
\label{37ex}
\end{center}
\end{figure}

Assuming free composite fermions, the number of states in $\Lambda$ bands can thus be enumerated straightforwardly. 
Given the symmetry of the problem, these organize themselves into $L$ multiplets, with each multiplet consisting of $2L+1$ states.  The values of $L$ in a given band can be determined from the angular momentum algebra. For this purpose, we first list all configurations of fermions that correspond to a given band. A composite fermion in the lowest $\Lambda$ level has angular momentum $l=|Q^*|$; a composite fermion in the first excited LL has $l=|Q^*|+1$; a composite fermion in the second excited LL has $l=|Q^*|+2$; and so on. Given the occupations of different $\Lambda$ levels, we can determine all possible total angular momenta $L$ by a simple calculation. Filled bands can be left out in this calculation, as they do not contribute to the total angular momentum, and for bands that are almost full, it is easier to consider the CFHs rather than CFPs.  We must be careful here to take account of the fermionic nature of the CFPs or CFHs. Fig.~\ref{37ex} shows the various configurations for the first four bands at $\nu^*=3$ (i.e., $\nu=3/7$). 

For specificity, let us take the 3/7 state $(N,2Q)=(18,37)$ which maps into $(N,2Q^*)=(18,3)$ of composite fermions. We ask what is the spectrum of free composite fermions here. The ground state is unique, in which the lowest three $\Lambda$ levels, corresponding to angular momentum shells with $l=1.5$, 2.5 and 3.5, are full, producing $L=0$. For the first excited band, we create a CFP with $l_{\rm qp}=4.5$ and a CFH with $l_{\rm qh}=3.5$, which produces a single multiplet at each of $L=1, 2, \cdots, 8$. The second excited $\Lambda$ band consists of (i) a CFP-CFH pair with $l_{\rm qp}=5.5$ and $l_{\rm qh}=3.5$; (ii) a CFP-CFH pair with $l_{\rm qp}=4.5$ and $l_{\rm qh}=2.5$; or (iii) two CFPs with $l_{\rm qp}=4.5$ and two CFHs with $l_{\rm qh}=3.5$. The predicted number of $L$ multiplets is given in Fig.~\ref{spectra}c inside brackets.  

The analysis for other fractions is analogous. The number of multiplets for each $L$ in each band is shown in Figs.  \ref{spectra}, \ref{general} and \ref{second} inside brackets. We make the following observations:

\begin{itemize}

\item For the lowest $\Lambda$ band, the structure predicted by the free CF theory is in perfect agreement with that seen in exact diagonalization. That has been confirmed for a large many other configurations, and, to date, there is no known exception to this correspondence in the range of filling factors in the lowest LL where FQHE is seen. (We note that the correspondence does not always hold in the second LL, or at $\nu<1/5$ in the lowest LL where the crystal phase becomes relevant.)

\item For higher $\Lambda$ bands, the predicted $L$ values in general do not agree with the observed ones. The disagreement is minimal for the first excited band at the Jain fractions $n/(2pn\pm 1)$, i.e.the single exciton branch, where the only deviation is that the free-CF counting predicts one multiplet at $L=1$ whereas no $L=1$ multiplet is observed in the exact spectra. For some cases, the deviation can be much larger. For example, for $(N, 2Q)=(16,32)$, there are 16 $L=5$  multiplets, whereas the correspondence with the noninteracting system $(N, 2Q^*)=(16,1)$ predicts 27 $L=5$ multiplets. 

\item The number of multiplets predicted by the free CF theory is never less than that observed in numerical experiments. The free CF theory thus over-predicts the number of excited states.

\end{itemize}

This illustrates the puzzle of the missing states, an explanation of which is the goal of the present article. We note that it is clear that the mapping between free composite fermions and the interacting electrons in the lowest LL {\em must} break down at some energy, because, for a system with finite $N$, the FQHE system in the lowest LL has a finite number of linearly independent states, whereas the free fermion system with an infinite number of Landau levels has an infinite number of linearly independent states. However, this observation by itself does not explain why the mapping breaks down already at the first excited band; there are certainly sufficiently many states available in the lowest Landau band to allow the mapping to hold for several lowest bands. A possibility, which would {\em a priori} seem more likely, would be that the mapping would continue to hold, as we go to higher and higher bands, until all lowest LL states are exhausted.  It is also unclear why the number of missing states depends on the filling factor and the angular momentum $L$ in a seemingly haphazard fashion as seen above.

With more work, the CF theory actually explains the excited bands as well. While we will be concentrating on the counting rules, the CF theory also tells us how to construct explicit wave functions for the $\Lambda$ band states at $2Q$ starting from the Landau band states at $2Q^*$.  In this method, we first construct all wave functions in the $q^{\rm th}$ Landau band of $N$ electrons at $2Q^*$, and denote this basis by $\{\Phi_{N,2Q^*,q,L,\alpha}\}$, where we have chosen the basis functions to have definite total orbital angular momentum $L$, and $\alpha$ labels the different orthogonal wave functions with the same $L$ quantum numbers. Now we construct Jain's CF wave function corresponding to each basis function as (in Haldane's spherical geometry \cite{Haldane1983})
\begin{equation}
\Psi^{CF}_{N,2Q,q,L,\alpha} = {\cal P}_{\rm LLL} \prod_{j<k}(u_iv_j-v_iu_j)^{2p}\Phi_{N,2Q^*,q,L,\alpha} 
\end{equation}
which produces a correlated basis $\{ \Psi^{CF}_{N,2Q,q,L,\alpha} \}$ at $2Q=2Q^*+2p(N-1)$. Here $u=\cos(\theta/2)e^{\phi/2}$ and $v=\sin(\theta/2)e^{-i\phi/2}$ denote the position of a particle on the sphere ($\theta$ and $\phi$ are the angular coordinates); $q$ labels the $\Lambda$ bands; and it can be proven that the angular momentum quantum number $L$ remains invariant under this mapping \cite{Jain_book}. The symbol ${\cal P}_{\rm LLL}$ denotes lowest LL projection, which is accomplished either by the Dev-Jain method \cite{DevJainPRL92} (which can be implemented for small systems with, typically, $N\leq 10$) or by the Jain-Kamilla method \cite{JainKamilaIJMPB1997} (which enables a treatment of large systems as well). We then orthogonalize this basis, and diagonalize the {\em full} Coulomb Hamiltonian in this basis. This program is called ``CF diagonalization" (CFD). For small systems, orthogonalization can be performed by expressing the wave functions in the Slater determinant basis; for larger systems, a Monte Carlo method has been developed by Mandal and Jain  \cite{Mandal02} to perform CFD (the evaluation of various multi-dimensional matrix elements is performed by the Monte Carlo method, in which a desired accuracy can be obtained by running the computer for a sufficiently long time). The various steps are complicated, but have been given in the literature \cite{JainKamilaIJMPB1997,Mandal02} and will not be repeated here, except to note that no approximations are made. 
 
A remarkable observation made by Wu and Jain \cite{WuJain95} (also see Ref.~\onlinecite{DevJainPRL92}) was:

\begin{enumerate}

\item When we composite-fermionize all (orthogonal) states of a Landau band at $2Q^*$, the resulting states at $2Q$ are often not all linearly independent. The resulting linearly independent basis at $2Q$ is thus, in general, smaller than that at $2Q^*$.  Such linear dependences are surprising because the actual wave functions are very complex, and indicate some hidden mathematical structures in Jain's CF wave functions that are yet to be discovered.

\item The number of linearly independent states at $2Q$ agrees with that seen in exact spectra for all cases studied so far. 

\end{enumerate}

We have found this to be valid for all additional cases studied during the course of the present work. This tells us that the CF theory allows, in principle, a calculation of the number of states in each band. Nevertheless, the way things stand at this stage, to get the correct counting it is necessary to construct the orthogonal CF basis, which requires a complex computer calculation. That is unsatisfying, because: it can be accomplished only for small systems; it does not give us an insight into the origin of the missing states; and it fails to clarify if the missing states are thermodynamically significant. Our aim in this article is to bring further clarity to this issue.

\section{Exclusion rules for single excitons}

Having shown that the model of noninteracting composite fermions produces spurious unphysical states in the excitation spectrum, it is natural to ask if the excited $\Lambda$ bands can be explained in terms of {\em interacting} composite fermions. This interaction must be infinitely strong, so that it will eliminate some states by sending them to infinity. Further, it must do so in precisely such a manner as to bring the ``modified Landau bands" into one-to-one correspondence with the $\Lambda$ bands. It will also be most satisfying, and also meaningful, if the interaction is a relatively simple 2-body interaction.  

How do we identify the form of the interaction that will accomplish the task? 
Let us first note that no spurious states appear in the lowest Landau band, which implies that no interaction is needed for fermions in the ``ground" manifold. (Please recall that we are concerned here with a strong interaction that would eliminate states; there is of course a weak residual interaction between composite fermions that would broaden the ground manifold, but without altering the number of states.) For the incompressible states at 1/3, 2/5 and 3/7 (upper panels) the $L=1$ exciton is missing in the first excited $\Lambda$ band. This suggests that we should impose an interaction that eliminates the $L=1$ CF exciton. As noted above, the {\em microscopic} CF theory actually predicts this elimination, because a ``composite-fermionization" of the $L=1$ exciton wave function at $\nu^*=1$, 2 or 3 annihilates it \cite{DevJainPRL92} (also see Ref.~\onlinecite{HeSimonHalperin94}). 

\begin{figure}[b]
\includegraphics[width=0.4\textwidth]{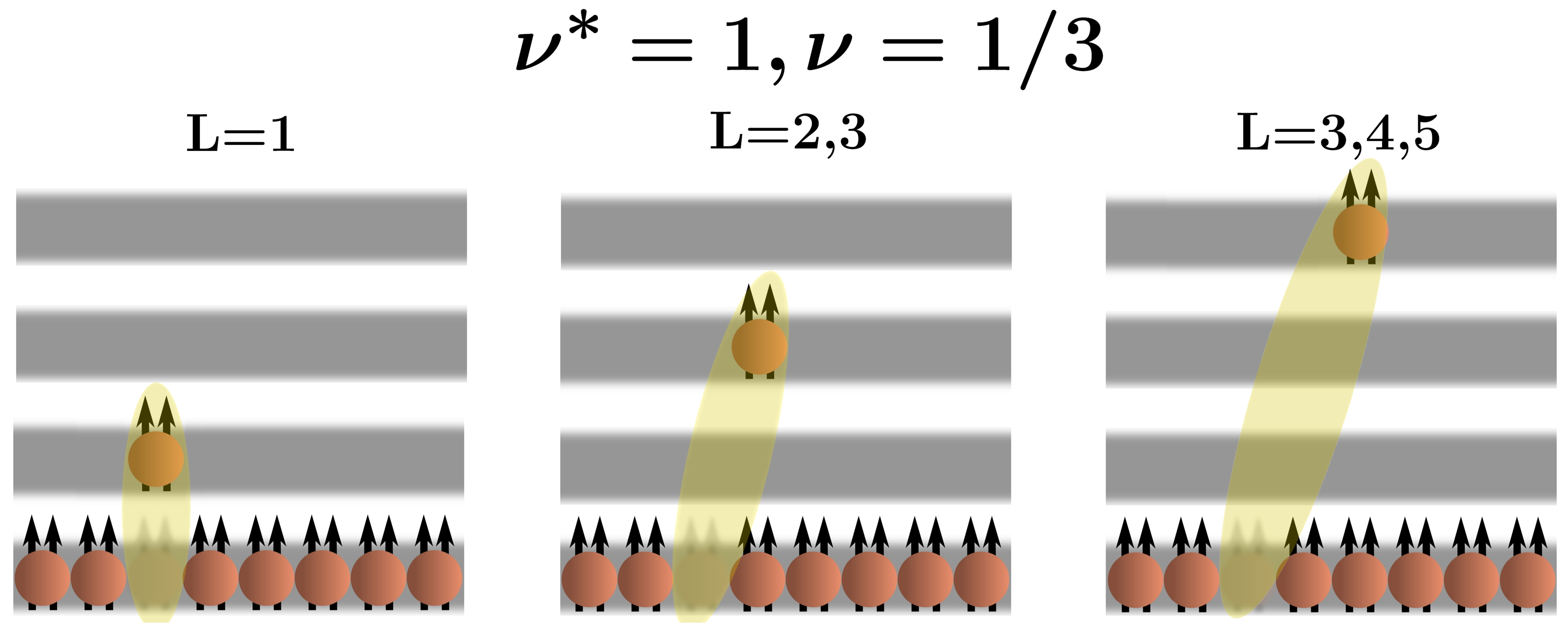}
\includegraphics[width=0.4\textwidth]{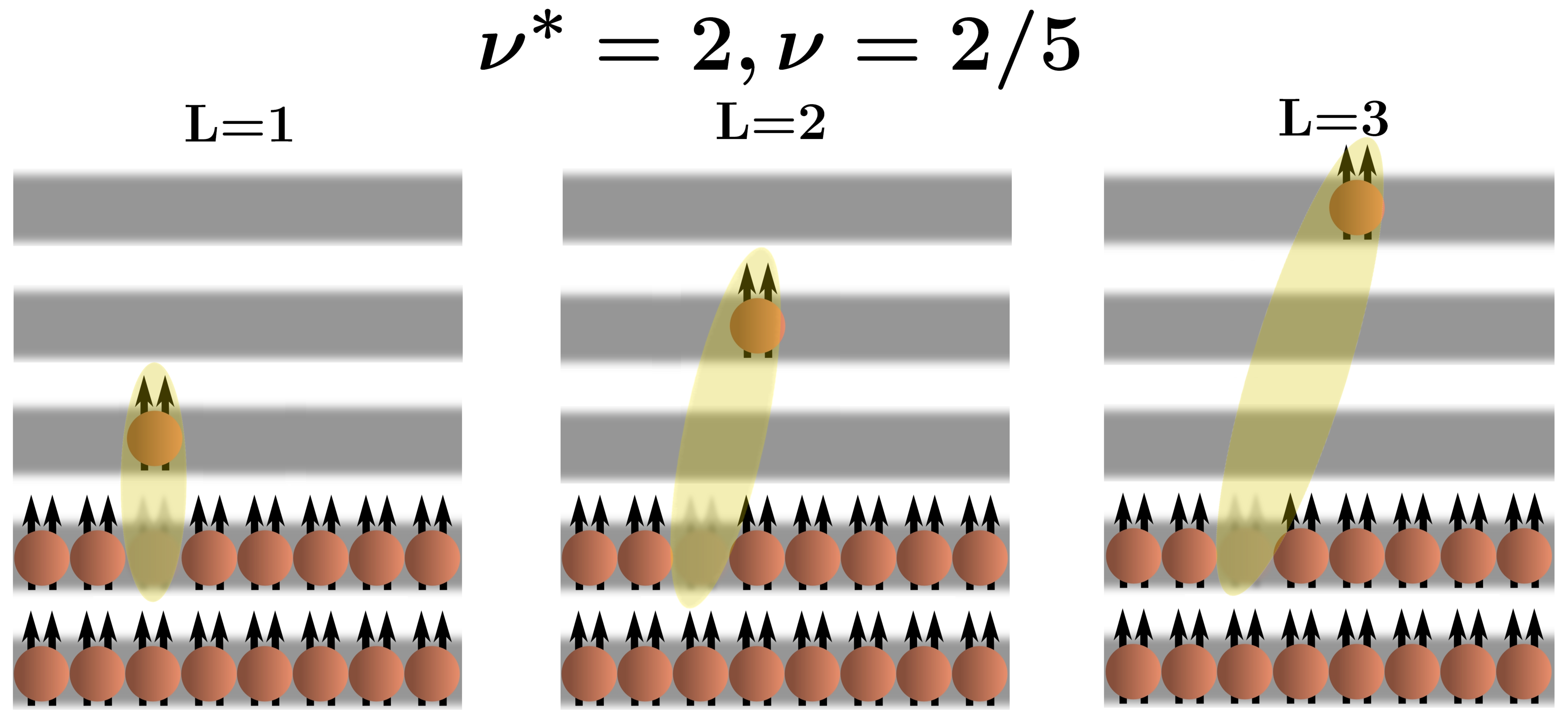}
\includegraphics[width=0.4\textwidth]{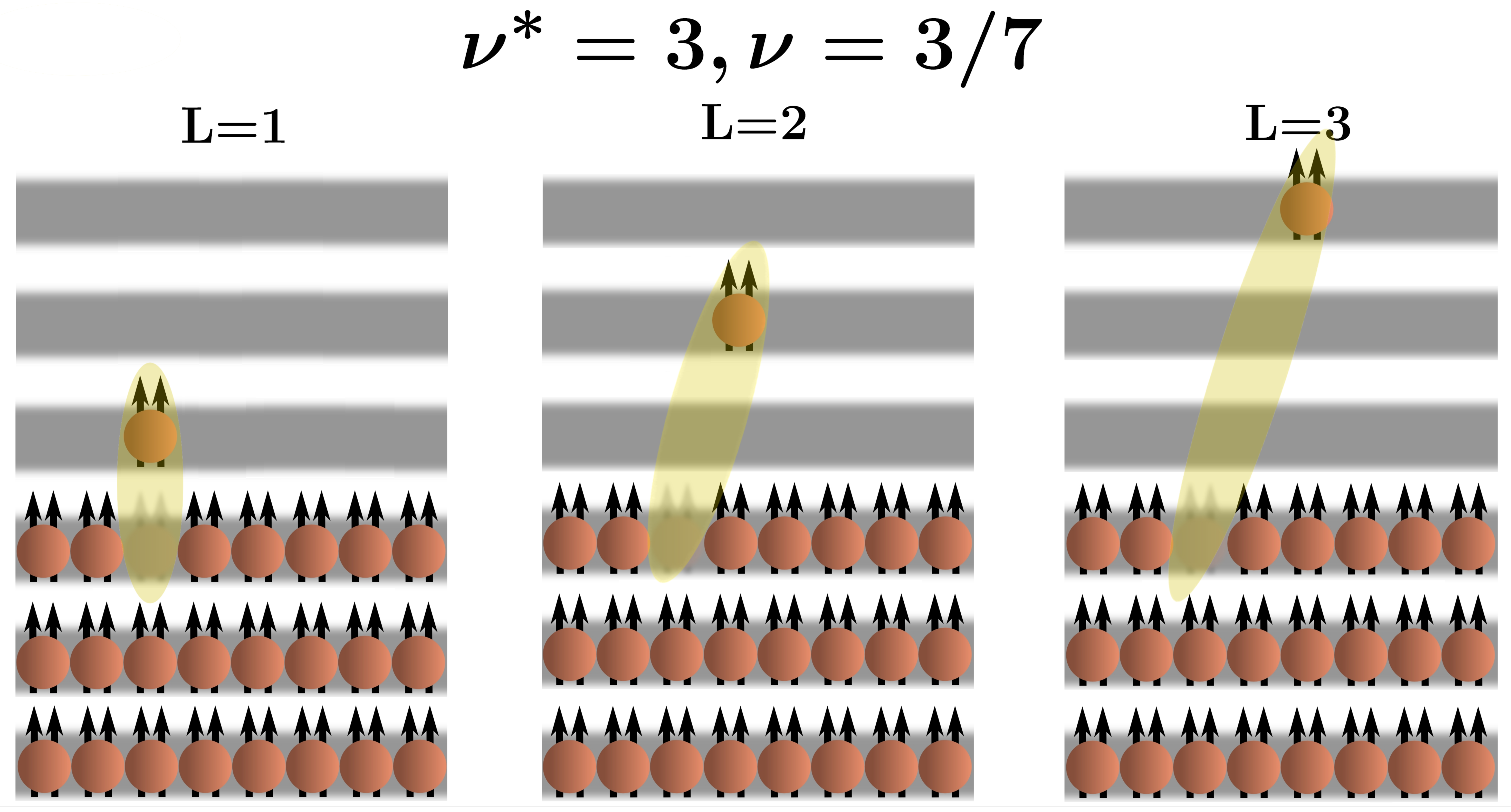}
\caption{This figure depicts the missing CF excitons (their angular momenta $L$ are shown) up to 3 units of ``CF cyclotron energy." The top, middle and bottom panels are for excitations out of one, two and three filled $\Lambda$ levels, respectively.}
\label{excl}
\end{figure}

We make the simplest assumption that the interaction acts only on particle-hole pairs of composite fermions. A justification of this assumption can come only {\em a posteriori}. If it does not work, or works only partially, we will need to return to this assumption and modify it. Given that the microscopic CF theory contains information about the missing states, we are motivated to use the CF theory itself to derive the constraints on single excitons.  We use the notation in which $|n\rightarrow n+k; L\rangle$ labels a single exciton at $L\geq k$ in which an electron has been excited from the $n^{\rm th}$ LL ($n=1, 2, \cdots$) by $k=1, 2, \cdots$ units of CF cyclotron energy $\hbar\omega_c^*$.  (No exciton with $L<k$ is possible for an exciton in which a composite fermion has been excited across $k$ $\Lambda$ levels because the angular momenta of the CFP and CFH differ by $k$.) We construct wave functions of single CF excitons with progressively higher cyclotron energies and ask how many are eliminated upon composite-fermionization. We have carried out this program for $k\leq 3$ and our findings are summarized in the following paragraph and in Fig. \ref{excl}.

We find that for the fractions $\nu=n/(2n+1)$ with $n= 2, 3$, the CF exciton $|n\rightarrow n+L; L\rangle$ is to be excluded because its wave function is linearly dependent on the CF exciton wave functions at the same $L$ in lower bands. (We note that there is flexibility as to which of the linearly dependent CF excitons we exclude; we have made choices which we find to be most convenient.) For $\nu=1/3$ all excitons $|1 \rightarrow 1+k; L \rangle$ with $L\leq 2k-1$ are eliminated. Thus the following CF excitons are excluded for $k\leq 3$ (shown pictorially in Fig.~\ref{excl}): 
\begin{eqnarray}
&\nu={1\over 3}:& |1\rightarrow 2; 1\rangle, |1\rightarrow 3; 2\rangle, |1\rightarrow 3; 3\rangle, \nonumber \\
& & |1\rightarrow 4; 3\rangle, |1\rightarrow 4; 4\rangle, |1\rightarrow 4; 5\rangle \nonumber \\
&\nu={2\over 5}:& |2\rightarrow 3; 1\rangle, |2\rightarrow 4; 2\rangle |2\rightarrow 5; 3\rangle \label{exclusion}\\
&\nu={3\over 7}:& |3\rightarrow 4; 1\rangle, |3\rightarrow 5; 2\rangle, |3\rightarrow 6; 3\rangle \nonumber
\end{eqnarray}
These exclusion rules have been demonstrated exactly (numerically) by performing CFD on all $N$ values up to 14, 16 and 18 for $\nu=1/3$, 2/5 and 3/7, respectively, and are expected to be valid for arbitrary $N$. We do not have a simple understanding of these exclusion rules or for the difference between the rules at 1/3 and those at other fractions of the form $n/(2n+1)$. Nonetheless, it should be stressed that these exclusion rules are not {\em ad hoc} but are a consequence of exact linear dependences of the CF exciton wave functions. This extends previous works in Refs. \onlinecite{DevJainPRL92,WuJain95,Majumder09} on the elimination of single excitons.

In the following, we explore the consequences of these exclusion rules. These will surely modify the spectrum, but the question is to what extent the modified spectrum agrees with that seen in exact diagonalization studies.

\section{Modified spectrum}

We now proceed to construct the modified $\Lambda$ bands in which the above excitons are excluded. For a general case, we find it convenient to diagonalize the basis $\{\Phi_{N,2Q^*,q,L,\alpha}\}$ for an interaction which imposes a very heavy penalty for the excitons enumerated above, and retain only the zero energy states.  (To obtain $L$ eigenstates, we also add other pseudopotentials with a very small coefficient of order $10^{-6}$. This interaction is small enough that it allows us clearly to identify the zero energy states.) Take for example the second excited band of the states with $n\geq 2$ filled $\Lambda$ levels. This contains two kinds of states. In one, a single CFP is excited across two $\Lambda$ levels; we already know, from our rules above, which of these are to be eliminated. The nontrivial part consists of the two CF excitons, each produced by exciting a single particle across one $\Lambda$ level.  To project out all states containing the $|n\rightarrow n+1;1\rangle$ exciton, we diagonalize our interaction in the basis of two particles and two holes with appropriate angular momenta and identify the zero energy states.  Combining the results produces the number of multiplets shown outside brackets in the top panels of Fig.~\ref{spectra}. All other systems can be treated analogously, and Figs.~\ref{spectra}, \ref{general} and \ref{second} show the number of remaining $L$ multiplets in each modified Landau band.

For some simple cases, the counting can be done straightforwardly, as we illustrate by taking a couple of examples. It is convenient directly to count the number of eliminated states. First consider the second excited band of the $\nu^*=3$ system with $(N,2Q^*)=(18,3)$ [i.e. $(N,2Q)=(18,37)$]. Here, there are two kinds of states, those containing two excitons and those containing a single exciton. In the first kind, we wish to eliminate all configurations in which one of the two excitons has $L=1$.  We therefore ask how many multiplets can be formed by combining two excitons, with one exciton at $L=1$ and the other at $L=1, 2, \cdots, 8$. We use the relations:
\begin{eqnarray}
1 \otimes 1 &=& 0 \oplus 2 \nonumber\\
1 \otimes 2 &=& 1 \oplus 2 \oplus 3 \nonumber\\
1 \otimes 3 &=& 2 \oplus 3 \oplus 4 \nonumber\\
& \cdots & \nonumber
\end{eqnarray}
(In the first equation above the two excitons are identical bosons, which is the reason why $L=1$ is missing on the right hand side; for all other cases, the two excitons have different angular momenta and hence are distinguishable particles.) This predicts $1, 1, 3, 3, 3, 3, 3, 3, 2, 1$ missing multiplets at $L=0, 1, \cdots, 9$. In the second kind of states with a single exciton (in which a single composite fermion has been excited across two $\Lambda$ levels), the $L=2$ multiplet is missing as per our exclusion rules. Together, this produces $1, 1, 4, 3, 3, 3, 3, 3, 2, 1$ missing multiplets at $L=0, 1, \cdots, 9$, which are precisely the number of eliminated multiplets observed in exact diagonalization (i.e., the difference between the numbers inside and outside the brackets in Fig.~\ref{spectra}c for the second excited band). For the system in Fig.~\ref{spectra}f that contains a CFP in the ground state, we add an $L=1$ exciton to a CFP with $L=4$ to predict a single missing multiplet at each of $L=3, 4, 5$. For the the  system in Fig.~\ref{spectra}f which contains a CFH in the ground state, we add an $L=1$ exciton to a CFH with $L=3.5$ to obtain a single missing multiplet at each of $L=2.5, 3.5, 4.5$. These also agree perfectly with the correct counting. The analysis for the other panels can be performed analogously.

Our calculations show that for $\nu^*>1$ ($\nu>1/3$), the elimination of the above mentioned excitons brings the counting of states into perfect agreement with the exact results.  This is true even of the second excited band, wherever it is identifiable in the exact spectra. We have considered all $N$ values accessible to our exact diagonalization study (up to 16 and 18 particles for 2/5 and 3/7, respectively), and confirmed the agreement whenever well defined $\Lambda$ bands can be identified.

For $\nu^*\leq 1$ ($\nu\leq 1/3$), the modified counting is in much better, but not perfect, agreement with the exact results. There is slight mismatch for the second excited band at $\nu=1/3$ (Fig.\ref{spectra}a) and for the first excited band of the CFH of $\nu=1/3$ (Fig.~\ref{spectra}g); the predicted numbers that do not match the exact ones are shown by green color. (The exact number can be obtained by counting the number of multiplets in the spectrum.) The predicted number is still slightly greater than that observed, which possibly indicates the need for an additional three-body constraints, e.g. on the 2CFH+1CFP trion state at $\nu^*=1$. We have not pursued this further.

\section{Quantitative confirmation}

As mentioned above, we can go beyond the counting of multiplets of various $\Lambda$ bands to construct explicit basis functions for them by  CFD. As also mentioned above, composite-fermionization of the $\Lambda$ bands of {\em free} composite fermions produces the correct counting; in this case, the number of linearly independent states {\em after} composite-fermionization is less than that before, and precisely equal to the number of physical states in the $\Lambda$ bands. We have confirmed that composite-fermionization of the modified spectrum does not eliminate any states. This shows how we can identify the surviving states {\em without} actually doing the full CFD.

\begin{figure}[t]
\includegraphics[width=0.48\textwidth, height=0.3\textwidth]{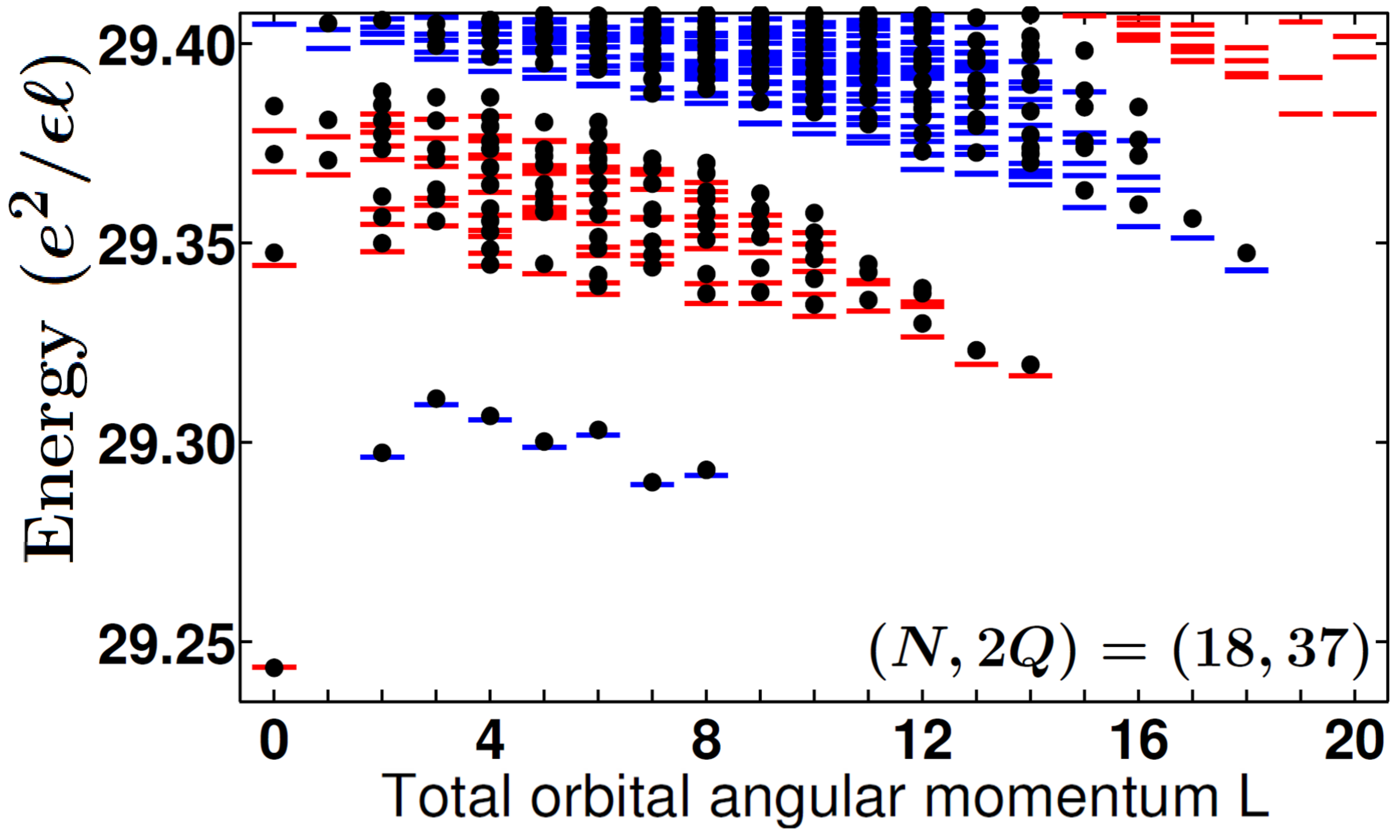}
\caption{CF spectra for $(N,2Q)=(18,37)$ obtained by performing CF Diagonalization in the basis derived from all states up to the third excited Landau band. This corresponds to $\nu^*=3$ ($\nu=3/7$) and should be compared with Fig. \ref{spectra}c which includes all states up to the second excited band.}
\label{Figure3}
\end{figure}

In either case, CFD produces a detailed prediction for the actual spectra. The spectra obtained from CFD are shown by dots in Fig.~\ref{spectra}, {\em quantitatively} establishing the validity of the CF description for the excited $\Lambda$ bands.  While the agreement for the excited $\Lambda$ bands is less accurate than that for the lower $\Lambda$ bands, it is still excellent and captures the energy splittings very nicely. The slight discrepancy arises because in the actual system, composite fermions mix with higher $\Lambda$ level states to lower their energy. Such mixing is small, as indicated by the smallness of the deviation between the exact and the CFD energies. However, if need be, more accurate energies may be obtained from the CF theory by allowing $\Lambda$L mixing by incorporating yet higher $\Lambda$ bands in the CFD calculation. An explicit example is shown in Figure \ref{Figure3} for $\nu=3/7$, which gives a much more accurate account of the second and third excited $\Lambda$ bands that than that in Fig.~\ref{spectra}c .

As we mentioned above, in addition to the infinite hard-core interaction that eliminates certain excitons, there exists a weak residual interaction between composite fermions that is responsible for a splitting of the states in a given $\Lambda$ band. This is best captured by the process of CFD, which produces the splittings very accurately. One may ask if there is a simple effective interaction between CFP-CFP, CFP-CFH and CFH-CFH that will describe that splitting \cite{Sitko,Lee1}. That question is outside the scope of this paper.

\section{Concluding remarks}

In summary, we have shown that the entire many particle excitation spectrum of the FQHE can be understood in terms of a model of {\em interacting} composite fermions. There are two kinds of interaction between composite fermions: (i) A strong interaction that eliminates some states by pushing them to infinity; this acts as a constraint on the allowed space of states. (ii) A weak interaction that removes the degeneracy of the remaining states and results in a broadening of the $\Lambda$ bands. The second interaction can be treated accurately by the method of composite-fermion diagonalization. Our main accomplishment in this article is to identify the strong interaction and show how it produces a correct counting of states in various $\Lambda$ levels. 

It is remarkable that such an excellent and detailed understanding is possible for a strongly correlated system with inherently non-perturbative physics. We close with a few observations.

The strong interaction has a strange form, in that it contains no CFP-CFP or CFH-CFH term; it only acts between an excited composite fermion (CFP) and the hole left behind (CFH). Also, while the interaction has a 2-body form for composite fermions, it represents very complex many-particle structures of electrons. To begin with, composite fermions and their $\Lambda$ levels are themselves complicated, emergent entities. The CF hole represents a collective state of composite fermions in which a composite fermion is missing in an otherwise full $\Lambda$ level.

Our work has implications for experiments. It confirms in detail the reality of higher unoccupied $\Lambda$Ls, thus providing support for the identification of the high energy modes observed in by Rhone {\em et al.} \cite{Rhone11} as single CF excitons across more than one $\Lambda$L \cite{Rhone11,Majumder09}. It is noted that this experiment measures modes at relatively large wave vectors (i.e., large $L$), whereas the eliminated CF excitons have small wave vectors (i.e. small $L$). 

The following two observations shed light on why FQHE states along $n/(2n+1)$ with high $n$ are so stable, and why the CF theory has proved successful for the CF Fermi sea. (i) A remarkable feature emerging from the exact spectra is that the $\Lambda$ bands become better defined for $n/(2n+1)$ states with higher $n$. For example, at $\nu^*=1$ the second excited band merges with the third excited band for $N>6$. In contrast, the second excited band at $\nu^*=3$ is sharply defined even for $N=18$. This demonstrates that the effective interaction between composite fermions becomes weaker as their filling factor is increased.  This is surprising, because one expects that as the gaps get smaller, there would be a stronger mixing between different $\Lambda$ levels. Our results show that the residual interaction between composite fermions (which is what causes broadening of a $\Lambda$ band) decreases faster than the CF cyclotron gaps separating the $\Lambda$ levels as we increase $n$.  Some insight into this behavior can be gained from the observation that the residual interaction between composite fermions is proportional to $e^{*2}=(2n+1)^{-2}$, where $e^*$ is magnitude of the local charge of the CFP or the CFH, whereas the CF cyclotron energy is proportional \cite{HalperinLeeRead1993} to $\hbar\omega_c^*=eB^*/m^*c=eB/(2n+1)m^*c$, where $m^*$ is the CF mass and $B^*$ is the effective magnetic field sensed by composite fermions. (ii) We can further deduce from our results that the number of eliminated states is thermodynamically insignificant, at least at low temperatures when the density of excited CF particles and CF holes is small. The reason is that only those states are eliminated in which a CFP and a CFH are close to one another, which is not a probable configuration at sufficiently small densities. (We note that this conclusion requires the understanding of the missing states developed above. Otherwise, in small system studies, it appears as though a large fraction of states is missing.) It can also be seen that the exclusion rules do not affect the counting of states at large momenta, which is where most states appear (including the $2L+1$ multiplicity), and which are also the states probed by certain experiments. These observations altogether offer insight into why the analysis in terms of nearly free composite fermions has proved so successful for the physics of the CF Fermi sea. 

We expect similar physics at the Jain fractions $n/(2pn-1)$, but have not pursued that in this work becaue these fractions involve reverse flux attachment \cite{WuDevJain}, which is technically more difficult to deal with. It would be natural to seek a generalization of the above considerations to include spin; the situation is likely be more complicated there because the Hund's rule for composite fermions already eliminates some of the states predicted by the model of free composite fermions \cite{WuJain96}. It would also be interesting to explore the origin of missing states within the Chern-Simons \cite{HalperinLeeRead1993,Lopez}, Hamiltonian \cite{MS}, and the conformal field theory \cite{Hansson} treatments of composite fermions.

\section{Acknowledgments}

We are grateful to S.S. Mandal for many useful discussions and for help with computation.  We acknowledge financial support from the NSF grants DMR-1005536 and DMR-0820404 (Penn State MRSEC), the Polish NCN grant 2011/01/B/ST3/04504 and the EU Marie Curie Grant PCIG09-GA-2011-294186 (AW). We thank Research Computing and Cyberinfrastructure, a unit of Information Technology Services at Pennsylvania State University, as well as Wroclaw Centre for Networking and Supercomputing and Academic Computer Centre CYFRONET, both parts of PL-Grid Infrastructure.

\end{document}